\documentclass[12pt]{article}
\usepackage{amsfonts}
\usepackage{latexsym,amsmath}
\usepackage{amssymb,array,graphicx}
\begin{document}
\title{An application of time truncated single acceptance sampling inspection plan based on transmuted Rayleigh distribution}
\author{Harsh Tripathi $^{a}$\footnote{Corresponding author. e-mail: 2017phdsta01@curaj.ac.in} and Mahendra Saha $^{a}$\\
\small $^{a}$Department of Statistics, Central University of Rajasthan, Rajasthan, India}
\date{}
\maketitle
\begin{abstract}
In this paper, we introduce single acceptance sampling inspection plan (SASIP) for transmuted Rayleigh (TR) distribution when the lifetime experiment is truncated at a prefixed time. Establish the proposed plan for different choices of confidence level, acceptance number and ratio of true mean lifetime to specified mean lifetime. Minimum sample size necessary to ensure a certain specified lifetime is obtained. Operating characteristic(OC) values and producer's risk of proposed plan are presented. Two real life example has been presented to show the applicability of proposed SASIP.
\end{abstract}
\section{Introduction}
In a global market, product quality is essential for the sustenance of any business, industry or enterprises. There are two important techniques for ensuring product quality: the statistical process control which is achieved through the techniques of control charts and process capability analysis and the statistical product control which is achieved through the techniques of acceptance sampling inspection plan(ASIP). ASIP is an essential tool in the Statistical Quality Control (SQC). ASIP is helpful to maintain the quality of products in which both producers and consumers are interested.  ASIP is a product control technique carried out in industrial sectors to inspect the quality of the product of a sample taken from the lot against the specified quality standards before delivering the product to the market. Besides minimizing the cost and time required for the quality control or reliability tests, acceptance sampling technique helps to decide about the acceptance or rejection of the submitted lot of products and thus provides the desired protection to producers and consumers. There are a number of different ways to classify acceptance sampling plans. One of major classification is by variables and attributes [see, Montgomery et al. ($2011$)]. Again, attribute sampling plan may be of Single acceptance sampling inspection plan(SASIP), double acceptance sampling inspection plan (DASIP), multiple acceptance sampling inspection plan (MASIP), sequential sampling plan, and skip-lot sampling plan etc., while variable sampling plan uses the accurate measurements of quality characteristics for decision-making rather than classifying products as conforming (or non-defective) or non-conforming (or defective) in attribute sampling plans. SASIP is mostly used in practice because this is simple to use, and so the main focus of this paper is on the kind of acceptance sampling plans under time truncated censoring scheme. In SASIP, the decision is made on the basis of one single sample selected from the lot.\\

In time truncated sampling inspection plan, we consider time as an important factor to set the quality of products or lot. Time truncated sampling plan is used mostly when testing is destructive, time consuming and cost consuming. In time truncated scheme terminate the life test at pre-determined time t or cumulative number of failures larger than the given acceptance number $c$ at prefixed time $t$, the lot is accepted. Otherwise, if number of failure exceeds $c$, one can terminate the test before the time t and reject the lot. Thus, a truncated life test with a given an acceptance number $c$ could be conducted to find the smallest sample size to ensure a certain mean life of products with a specified consumer's confidence level $P^{*}$ (called the probability to control consumer's risk). Several authors have studied on time truncated SASIP for various models, viz., Epstein($1954$) introduced acceptance sampling plan based on time truncated life test in exponential distribution, Gupta and Groll($1961$) introduced acceptance sampling plan for gamma distribution, Gupta($1962$) considered time truncated acceptance sampling plan for log-normal distribution. In recent era, Baklizi et al. ($2004$), Rosaiah et al.($2005$), Balakrishnan et al.($2007$), Aslam et al.($2010$), Al-Nasser et al.($2013$), Rao et al.($2013$), Giu et al.($2014$) and Al-omari($2015$) have studied on time truncated SASIP for Birnbaum saunders distribution, inverse Rayleigh distribution, generalized Birnbaum saunders distribution, generalized exponential distribution, exponentiated frechet distribution, Marshall olkin extended exponential distribution, Gompertz distribution and generalized inverted exponential distribution respectively. Recently, Tripathi et al. ($2020$) and Saha et al. ($2021$) have developed SASIP for generalized half-normal distribution and transmuted Rayleigh distribution (zero one failure scheme) respectively.\\

The aim of this article to develop time truncated SASIP when lifetime of product follows the transmuted Rayleigh(TR) distribution where the shape parameter is known to us. The time truncated test is popularly used for a life test to save the time and cost required and is quite useful to take a decision regarding acceptance or rejection of lot. Table of minimum sample size has presented for which assures that specified mean life time is at least true mean life. The probability of acceptance and the producer's risk tables have presented also.\\

Rest of the article organized as follows: In Section $2$, we describe the probability density function (PDF) and cumulative distribution function (CDF) of TR distribution. In Section $3$, we show the structure of time truncated SASIP along with two subsections, namely, OC function and producer's risk of the sampling plan. In section $4$, we have discussed the tables, examples and figures. In Section $5$, we have presented two real life example to illustrate the application of proposed time truncated SASIP under TR distribution. Finally, conclusions about the article are given in Section $6$.

\section{Transmuted Rayleigh distribution}
Transmuted Rayleigh(TR) distribution has been introduced by Merovci ($2013)$. He derived some statistical properties such as moments, quantile function, order statistics of this distribution. Further, the model parameters are obtained by the method of maximum likelihood. Subsequently, Dey et al.($2017$) studied this distribution in greater details. They derived quantiles, moments, moment generating function, conditional moments, hazard rate, mean residual lifetime, mean past lifetime, mean deviation about mean and median, stochastic ordering, various entropies, stress-strength parameter and order statistics. Further, model parameters are obtained by different method of estimation, viz., the method of maximum likelihood, methods of moments, methods of L-moment, method of percentile, method of least squares, method of maximum product spacing, method of Cramer-von Mises, method of Anderson-Darling and right tail Anderson-Darling. This distribution has uni modal probability density function(pdf) and increasing hazard rate function. In several situations, only increasing hazard rate are used or observed. For example, Koutras(2011) found that software degradation times have increasing hazard rate function; Lai(2013) investigated the optimum number of minimal repairs for systems under increasing hazard rates etc.\\

TR distribution is useful to analyze the lifetime data. It is a generalization of Rayleigh distribution. A random variable $X$ is said to have a TR distribution if its PDF and cumulative distribution function (CDF) are given as
\begin{eqnarray}\label{eq1}
f(x;\sigma,\lambda)=\frac{x}{\sigma^2}e^{\frac{-x^2}{2\sigma^2}}{(1-\lambda+2\lambda e^{\frac{-x^2}{2\sigma^2}})}~~~x>0,\sigma>0, |\lambda|\leq 1
\end{eqnarray}
\begin{eqnarray}\label{eq2}
F(x;\sigma,\lambda)={(1-e^{\frac{-x^2}{2\sigma^2}})}{(1+\lambda e^{\frac{-x^2}{2\sigma^2}})}
\end{eqnarray}
TR distribution have two parameters one is scale parameter $\sigma$ and second is transmuted parameter $\lambda$. If $\lambda=0$, then TR distribution converted into Rayleigh distribution. The r-th order raw moment, where X follows TR distribution is given below[see Merovci$(2013)$]
\begin{eqnarray}\label{eq3}
E(X^r)=\frac{1}{2} \sigma^r \Gamma(\frac{r}{2}) {(\lambda+2^{r/2}(1-\lambda))}; r=1,2,3.....
\end{eqnarray}
Mean of TR distribution has obtained by particular case r=1 in The r-th order raw moment and denoted by $\mu$ is given as
\begin{eqnarray}\label{eq4}
E(X)=\frac{\sigma \sqrt{\pi}}{2} {(\lambda+\sqrt{2}(1-\lambda))}=\mu
\end{eqnarray}
\section{Design of SASIP}
In this section, we describe the sampling plan when the quality of a lot follows TR distribution. We assume that the value of the transmuted parameter is known to us. Lot sentencing in SASIP depends on information provides by a single sample. SASIP consist following points of procedure:
\begin{itemize}
\item Select a random sample of size n and put them on test for given time t.
\item Let an acceptance number c, a lot is accepted if at most c failures out of n units occur within prefixed time t.
\item A ratio $t/\sigma_0$, where $\sigma_0$ is specified value of $\sigma$.
\end{itemize}
If the number of failures in a sample at most $c$, then lot is acceptable, otherwise reject the lot. We assume that the lot quality is represent by mean of the lot $\mu$. Larger lot mean shows better quality of the lot. $\mu$ is a function of $\sigma$ and $\mu_0$ be the specified mean lifetime of an item. Since the lot quality depends on $\sigma$, can be estimated by the mean of the TR distribution.\\
i.e., in particular for $\sigma=\sigma_0$
\begin{eqnarray*}
\sigma_0=\frac{2\mu_0}{\sqrt{\pi}} \frac{1}{{(\lambda+\sqrt{2}{(1-\lambda)})}}
\end{eqnarray*}
\begin{eqnarray*}
\mu_0=\sigma_0 \frac{\sqrt{\pi}}{2} {(\lambda+\sqrt{2}{(1-\lambda)})}
\end{eqnarray*}
A product is preferred by consumer, if the sample information support the hypothesis
$$
{H_0:\mu \ge \mu_0}
$$
it is clear that
$$
{H_0:\mu \ge \mu_0}\iff{H_0:\sigma \ge \sigma_0}
$$
Otherwise reject the lot of product. Consumer risk($\beta$) is fixed and not to exceed $(1-P^*)$, where $P^*$ be confidence level of consumer in sense of rejecting a lot of bad quality. It has mentioned that size of lot is very large so we can use binomial distribution. We want to find the smallest values of sample size $n$ for which observed number of failures within given time t does not exceed to acceptance number c, it is ensured that $\sigma\ge\sigma_0$ with minimum probability $P^*$. All the values of minimum sample size for different values of $P^*$, c, and $t/\sigma_0$ have presented in Table-$1$. 
\begin{eqnarray}\label{eq5}
\mbox{Minimize}~~~\mbox{ASN=n}\nonumber\\
\sum\limits_{i=0}^{c}{n \choose i} p^i (1-p)^{(n-i)}\le (1-P^*)
\end{eqnarray}
where, $p$ is the probability that an item fails before termination time, and for the considered model it is given as
\begin{eqnarray}\label{eq6}
p={\left(1-e^{\frac{-t^2}{2\sigma_0^2}\frac{\sigma_0^2}{\sigma^2}}\right)}{\left(1+\lambda e^{\frac{-t^2}{2\sigma_0^2}\frac{\sigma_0^2}{\sigma^2}}\right)}
\end{eqnarray}
$p$ depends on ratio of $t/\sigma_0$ for fix value of $\sigma=\sigma_0$. Minimum sample sizes satisfying Equation (\ref{eq5}) have been obtained for $P^*=0.75, 0.90, 0.95, 0.99$ and $t/\sigma_0=0.628,0.942,1.257,1.571,2.356,3.141,3.927,4.712$ for known value of $\lambda=0.5$. We have presented the values of minimum sample sizes $n$ in Table $1$.
\subsection{Operating characteristic function of sampling plan}
OC function of sampling plan provides the probability of acceptance of lot on basis of selected sample. For above plan probability is given below 
\begin{eqnarray}\label{eq7}
\mbox{PA}=\sum\limits_{i=0}^{c}{n \choose i} p^i (1-p)^{(n-i)}=P\{\mbox{accepting a lot}\}
\end{eqnarray}
OC values for considered plan has obtained by satisfying Equation \ref{eq7} for different values of minimum sample sizes and $\sigma/\sigma_0$. Probability of acceptance(OC values) has placed in Table-$2$ for fixed c=$2$ and $\lambda=0.5$.\\
\subsection{Producer's risk}
Producer's risk is described as probability of rejection of the lot when lot quality is good, i.e., $\mu \ge \mu_0$.\\
\begin{eqnarray*}
\mbox{PR}=1-P{(accepting ~a ~lot| \sigma\ge\sigma_0)}
\end{eqnarray*}
\begin{eqnarray}\label{eq8}
\mbox{PR}=\sum\limits_{i=c+1}^{n}{n \choose i} p^i (1-p)^{(n-i)}
\end{eqnarray}
For given value of producer risk, say $\delta$, we are interested to find the minimum ratio $\sigma/\sigma_0$ that will ensures the producer risk at most $\delta$. For sampling plan $(n,c,t/\sigma_0)$ and values of $P^*$, smallest value of $\sigma/\sigma_0$ satisfying Equation \ref{eq9}.
\begin{eqnarray}\label{eq9}
\sum\limits_{i=c+1}^{n}{n \choose i} p^i (1-p)^{(n-i)}\le\delta
\end{eqnarray}
Minimum value of ratio of $\sigma/\sigma_0$ have presented in Table $3$.

\section{Description and findings of tables, example and figures}
{\bf Tables:}\\
Table $1$ represents the minimum sample sizes required to ensure the true mean lifetime of product exceed than specified mean lifetime with probability at least $P^*$, corresponding acceptance number c and $\lambda=0.5$. From Table $1$, $P^*=0.95$, $t/\sigma_0=1.257$, $c=2$, the corresponding table value is $7$. When $7$ items put on test if not more than $2$ item fails before time t, then we accept the lot with probability 0.95.\\
$$
t/\sigma_0=1.257 \rightarrow t=1.257\sigma_0
$$
where $\sigma_0$ is specified value of $\sigma$ and estimated by specified mean of TR distribution.\\ 

In Table $2$, we provide the OC values for considered time truncated sampling plan for different values of $P^*$, $\sigma/\sigma_0$, $t/\sigma_0$ and $n$ at prefix value of $c=2$. Suppose $P^*=0.95$, $t/\sigma_0=1.257$, $\sigma/\sigma_0=4$ and $c=2$, the table value is $0.9898812$. This implies that lot is accepted, if $7$ items put on test if not more than 2 item fails before time t, then if true mean life is four times of specified mean lifetime, then the lot will be accepted with probability at least $0.9898812$.\\

Table $3$ consist the minimum ratio of  $\sigma/\sigma_0$ for acceptance of lot with producer risk $\delta=0.05$ and when $\lambda=0.5$. Values of Table $3$ shows that at what minimum value of $\sigma/\sigma_0$ producer risk is less than or eqaul to $0.05$. When $P^*=0.95$, $c=2$, $t/\sigma_0=1.257$, the table value of $\sigma/\sigma_0$ is $2.93$. It implies lot will be rejected with probability less than or equal to $0.05$. Values of producer's risk have placed in table $4$. Producer's risk has calculated when $P^*=0.95$, $c=2$ and $\lambda=0.5$.\\
All the following results have observed from given tables regarding the sample sizes, the probability of acceptance and producer's risk. Hence, all the tables have calculated for the given value of $t/\sigma_0$, $\sigma/\sigma_0$, c and $\lambda=0.5$.
\begin{enumerate}
\item From table $1$, we have observed that sample sizes n monotonically decreases when the value of the ratio $t/\sigma_0$ strictly increases corresponding to a particular value of c. Smallest value of sample size occurs when $t/\sigma_0=4.712$ for given value $c$. These result holds for all the set-ups of $P^*$  
\item From table $1$, we have observed the strictly increasing trend in sample size when acceptance number c is strictly increasing corresponding to a fix value of $t/\sigma_0$. Largest value of sample size occurs when $t/\sigma_0=0.628$ and acceptance number $c=10$. The Same pattern regarding sample sizes has observed for all the set-ups of $P^*$ 
\item From table $2$, we can see that largest OC(operating characteristic) values for a particular set-up of $P^*$ occurs when $t/\sigma_0=0.628$. This result holds for all the set-ups of $P^*$.
\item From table $2$, It has observed that OC(operating characteristic) values increases when the quality of product increases corresponding to a particular value of $n$ and $t/\sigma_0$. OC values increase rapidly towards to $1$ or nearer to $1$ when the value of true quality of lot(or product) is twelve times of the specified quality of lot(or product). These results hold for all the considered set-ups of $P^*$.
\item From table $4$, It can easily verify that producer's risk decreases when the quality of lot increases for a particular value of $t/\sigma_0$. In other words we can say that as much as the quality of lot increase than the chance of rejection of good lot decreases.
\end{enumerate}

{\bf Example:}\\
 Suppose producer wants to establish $1000$ units mean life time when the product under study follows TR distribution with $\lambda=0.5$. An experimenter wants to know the minimum sample size to be considered to make a decision about lot when true mean life time is $1000$ units with $P^*=0.95$ and $c=2$.
$$
\mu_0=1000
$$
$$
\sigma_0=935
$$
$$
t/\sigma_0=0.942 \rightarrow t=880
$$
If $11$ units put on test if not more than $2$ items fails before time $880$ units then accept the lot otherwise reject the lot.\\

{\bf Figures:}\\

This article consists three figures which show the graphical representation of findings of given tables and a brief discussion of all the figures are given below.\\
\begin{itemize}
\item {\bf Figure $1$} shows the trend of sample size against the ratio $t/\sigma_0$ for the fix value of $c=2$ and $P^*$. Largest value of sample sizes occurs for the largest confidence level $P^*$ i.e., $P^*=0.99$ and smallest value of sample sizes occurs for the smallest confidence level $P^*$ i.e., $P=0.75$. In wholesome, trend of sample sizes is increasing when $P^*$ increases.
\item {\bf Figure $2$} is a curve between the ratio $\sigma/\sigma_0$ and probability of acceptance for the set ups ($c=2, P^*=0.75$), ($c=2, P^*=0.90$), ($c=2, P^*=0.95$) and ($c=2, P^*=0.99$), corresponding to fix values of $t/\sigma_0$. Most of the times, there is decreasing trend of OC values when the ratio $t/\sigma_0$ is increasing for all the considered set ups of ($c, P^*$).
\item {\bf Figure $3$} is a curve between the ratio $t/\sigma_0$ and probability of acceptance for the set ups ($c=2, P^*=0.75$), ($c=2, P^*=0.90$), ($c=2, P^*=0.95$) and ($c=2, P^*=0.99$), corresponding to fix values of $\sigma/\sigma_0$. Most of the times, there is rapid increment in OC values when the ratio $\sigma/\sigma_0$ is increasing for all the considered set ups of ($c, P^*$).
\end{itemize}

\section{Real life examples}
In this section, we show the implementation of time truncated acceptance sampling inspection plan under TR distribution. For this purpose, we have consider two real life examples. First we check whether considered data set is well fitted with TR distribution or not. For this purpose we have considered two discrimination criteria based on the log-likelihood function evaluated at the maximum likelihood estimates of the parameters. These criteria are: AIC (Akaike Information Criterion), BIC (Bayesian Information Criterion) and are given by $AIC=-2l(\hat{\theta})+2p$, $BIC=-2 l(\hat{\theta})+ p\ln(n)$ respectively, where, $l(\hat{\theta})$ denotes the log-likelihood function evaluated at the MLEs, $p$ is the number of model parameters and $n$ is the sample size. Also, we provide the descriptive statistics, viz., minimum, $1$st quartile ($Q_1$), median, mean, $3$rd quartile ($Q_3$), maximum, coefficient of skewness ($CS$) and coefficient of kurtosis ($CK$) of the considered data sets in Table $5$. The model with lowest values for these statistics could be chosen as the best model to fit the data. The values of MLE of the parameter, $l(\hat{\theta})$, AIC, BIC and K-S statistic and corresponding $p$ are displayed in Table $6$, which we have done using $fitdistrplus$ package of $R$ software [see, Ikha and Gentleman ($1996$)] \\

\begin{itemize}
\item {\bf Data set I}, ordered failure times of release of software given in terms of hours from the starting of execution of software denoting the times at which the failure of software is experienced. Data can be regarded as an orderd sample of size 10 [see, Saha et al. (2021)]:
$$519,968,1430,1893,2490,3058,3625,4422,5218,5823$$
The values of  MLEs  of the parameters $\sigma$ and $\lambda$ are $2504.038$ and $0.1599$. The values AIC, BIC, K-S Statistic and P value of K-S statistics has been presented in table $6$. Now for our study we assume that experimenter wants to establish mean life is 1000 hours and also assume that $P^*=0.95$, $t/\sigma_0=0.628$ and $c=1$. Hence experiment time is $525$ hours. If number of failures before $525$ hours is less than or equal to $1$, we can accept the lot with mean $1000$ hours, with probability $0.95$. Since the number of failures before $525$ hours is $1$, therefore we can accept the lot with above specifications.

\item {\bf Data set II}, data set is given in Lawless 2003, and represents the number of millions revolutions to failure for 23 ball bearings.
$$17.88,28.92,33,41.52,42.12,45.60,48.40,51.84,51.96,54.12,55.56,67.80,68.64,68.64$$
$$68.88,84.12,93.12,98.64,105.12,105.84,127.92,128.04,173.40$$
The values of  MLEs  of the parameters $\sigma$ and $\lambda$ are $59.75319$ and $0.1594042$. The values AIC, BIC, K-S Statistic and P value of K-S statistics has been presented in table $6$. Now, we assume that experimenter wants to establish mean life is $50$ unit and also assume that $P^*=0.95$, $t/\sigma_0=0.628$ and $c=1$. Hence experiment time is $26$ unit. If number of failures before $26$ unit is less than or equal to $1$, we can accept the lot with mean $50$ hours, with probability $0.95$. Since the number of failures before $26$ unit is $1$, therefore we can accept the lot with above specifications.
\end{itemize}

\section{Conclusions}
In this article we developed a time truncated acceptance sampling inspection plan for transmuted Rayleigh(TR) distribution. We have presented table of minimum sample size which guarantee a certain mean life of units. We have also presented OC values of considered plan and minimum ratio of $\sigma/\sigma_0$ for known value of transmuted parameter $\lambda$. Proposed methodology can be used for other shape parameters. At the last of article we show the application of this methodology in real life scenario.

\section{References}
\begin{enumerate}
\item Epstein B (1953). Life testing. J Am Stat Assoc 48(263):486–502.

\item Gupta, S.S(1962). Life test sampling plans for normal and lognormal distributions,{\it Technometrics}, {\bf 4(2)}, 151-175.

\item Gupta, S.S. and Groll, P.A.(1961). Gamma distribution in acceptance sampling based on life test,{\it Journal of the American Statistical Association}, {\bf 56(296)}, 942-970.

\item Baklizi, A., EL Masri, A.E.K.(2004). Acceptance sampling plan based on truncated life tests in the Birnbaum Saunders model,{\it Risk Analysis}, {\bf 24}, 1453-1457.

\item Rosaiah, K. and Kantam, R.R.L.(2005). Acceptance sampling plan based on the inverse Rayleigh distribution,{\it Economic Quality Control,}{\bf 20(2)},77-286.

\item Balakrishnan, N., Lieiva, V., Lopez, J.(2007). Acceptance sampling plan from truncated life tests based on generalized Birnbaum Saunders distribution,{\it Communication in Statistics-Simulation and Computation}, {\bf 34(3)}, 799-809.

\item Aslam, M., Kundu, D., Ahmed, M.(2010). Time truncated acceptance sampling plans for generalized exponential distribution,{\it Journal of Applied Statistics}, {\bf 37(4)}, 555-566.

\item  Al-Nasser, A. D.,  Al-Omari, A. I. (2013). Acceptance sampling plan based on truncated life tests for exponentiated Frechet distribution. Journal of Statistics and Management Systems, 16(1), 13-24.

\item Rao, G.S., Ghitany, M.E., Kantam, R.R.L.(2008). Acceptance sampling plans for Marshall-Olkin Extended Lomax distribution,{\it International Journal of Applied Mathematics,}{\bf 21(2)}, 315-325.

\item Wenhao Gui1,2and Shangli Zhang. (2014).Acceptance Sampling Plans Based onTruncated Life Tests for Gompertz Distribution. Journal of Industrial Mathematics, http://dx.doi.org/10.1155/2014/391728.

\item A.I, Al-Omari.(2015). Time truncated acceptance sampling plans for generalized inverted exponential distribution,{\it Electronic Journal of Applied Statistical Analysis,}, {\bf 8(1)}, 1-12.

\item Tripathi, H., Saha, M., \& Alha, V. (2020). An Application of Time Truncated Single Acceptance Sampling Inspection Plan Based on Generalized Half-Normal Distribution. Annals of Data Science, 1-13.

\item Saha, M. Tripathi, H. and Dey, S. (2021). Single and double acceptance sampling plans for truncated life tests based on transmuted Rayleigh distribution. Journal of Industrial and Production Engineering, DOI: 10.1080/21681015.2021.1893843.

\item Merovci, F.(2013). Transmuted Rayleigh distribution,{\it Austrian Journal of Statistics}, {\bf 42(1)}, 21-31.

\item Dey S, Raheem E, Mukherjee S (2017). Statistical properties and different methods of estimation of transmuted Rayleigh distribution, {\it Revista Colombiana De Estadística}, 40(1):165–203.

\item Koutras VP. Two-level software rejuvenation model with increasing failure rate degradation. Depend Comput Syst. 2011;97:101–115.

\item Lai MT. Optimum number of minimal repairs for a system under increasing failure rate shock model with cumulative repair-cost limit. Int J Reliability Safety. 2013;7(2):95–107.

%\item  Al-Omari, A. I. (2014). Acceptance sampling plan based on truncated life tests for three parameter kappa distribution. Economic Quality Control, 29(1), 53-62.

\item Ihaka, R. and Gentleman, R. (1996). R: A language for data analysis and graphics. {\it Journal of Computational and Graphical Statistics}, {\bf 5}, 299-314.

\item J.F. Lawless (2003). Statistical Models and Methods for Lifetime Data,Wiley, New York.

%\item  Lio, Y. L., Tsai, T. R.,  Wu, S. J. (2009). Acceptance sampling plans from truncated life tests based on the Birnbaum–Saunders distribution for percentiles. Communications in Statistics-Simulation and Computation, 39(1), 119-136.
%
%\item  Lio, Y. L., Tsai, T. R.,  Wu, S. J. (2010). Acceptance sampling plans from truncated life tests based on the Burr type XII percentiles. Journal of the Chinese institute of Industrial Engineers, 27(4), 270-280.

%\item  Montgomery, D. C, Jennings, C. L, Pfund, M. E. (2011). Managing, Controlling and Improving Quality, New Jersey: John Wiley \& Sons.
%
%\item Sobel, M., Tichendrof, J.A.(1959). Acceptance sampling with new life test objectives,{\it Proceedings of the fifth National Symposium on Reliability and Quality Control,}{\bf Philadelphia},108-118.
%
%\item Tsai, T.R., Wu, S.J.(2006). Acceptance sampling plan based on truncated life tests for generalized Rayleigh distribution,{\it Journal of Applied Satistics,}{\bf 33}, 595-600.
%
%\item  Montgomery, D. C. (2009). Statistical Quality Control: A Modern Introduction, 6th Ed., New York: John Wiley \& Sons.

\end{enumerate}
%\newpage
\begin{figure}[htbp]
\resizebox{12cm}{8cm}{\includegraphics[trim=.01cm 2cm .01cm .01cm]{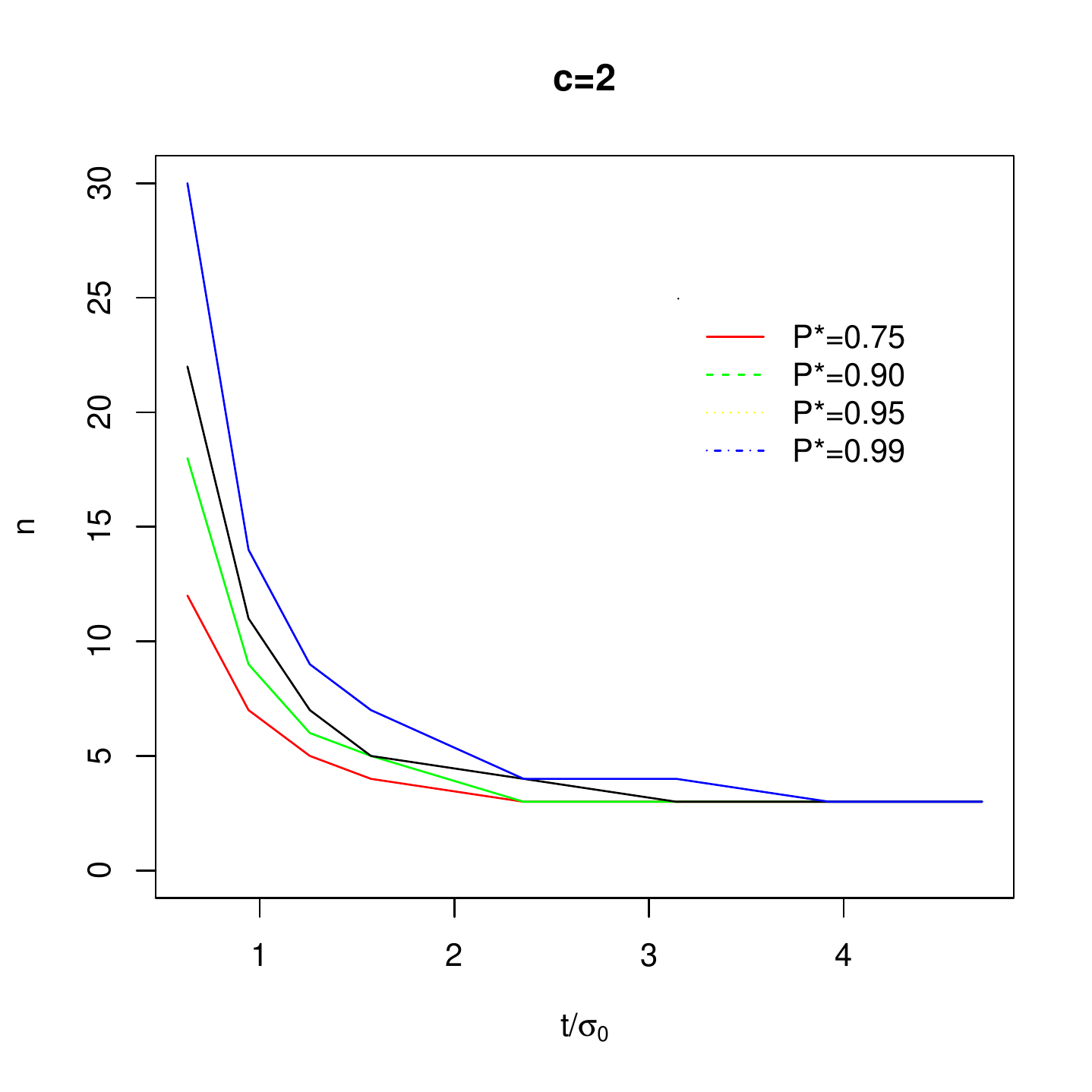}}
\vspace{1cm}
\caption{Ratio $t/\sigma_0$ verses sample size}
\label{fig1}
\end{figure}
\begin{figure}[htbp]
\centering
\resizebox{15cm}{8cm}{\includegraphics[trim=.01cm 2cm .01cm .01cm]{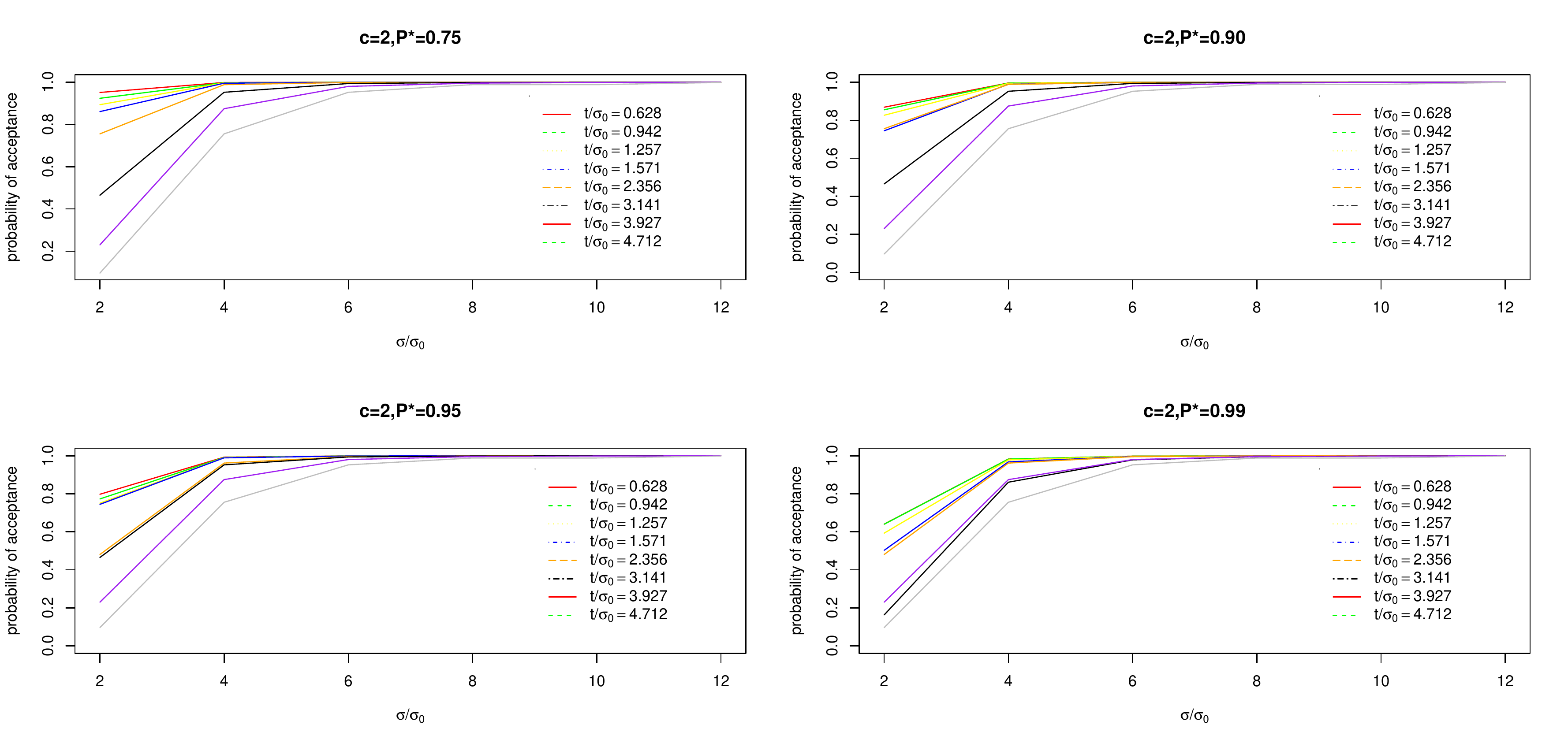}}
\vspace{1cm}
\caption{Ratio $\sigma/\sigma_0$ verses probability of acceptance curve}
\label{fig2}
\end{figure}

\begin{figure}[htbp]
\centering
\resizebox{15cm}{8cm}{\includegraphics[trim=.01cm 2cm .01cm .01cm]{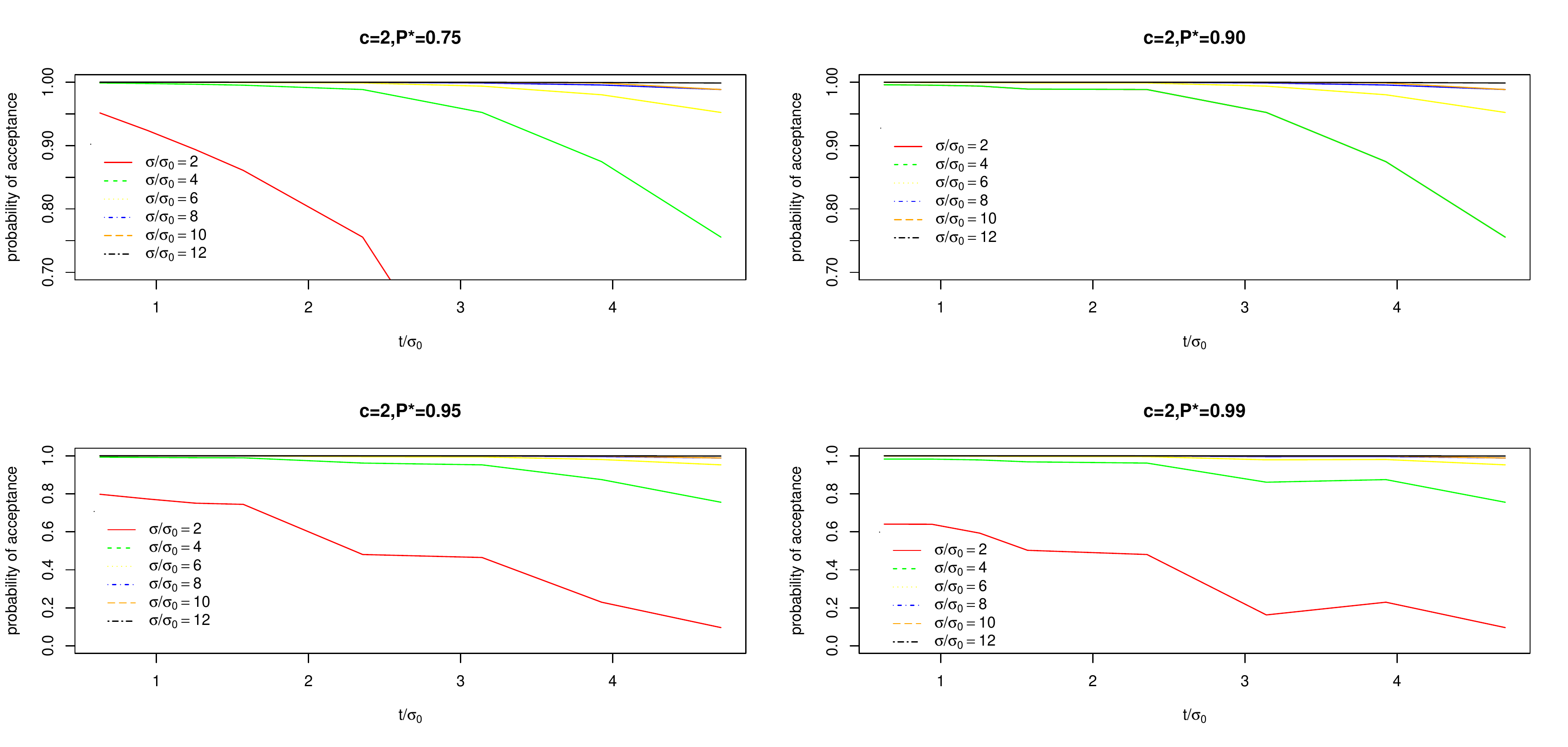}}
\vspace{1cm}
\caption{Ratio $t/\sigma_0$ verses probability of acceptance curve}
\label{fig3}
\end{figure}
%-------------------------------------------------------------------------------------------------
\begin{figure}[htbp]
\centering
\resizebox{15cm}{5cm}{\includegraphics[trim=.01cm 2cm .01cm .01cm]{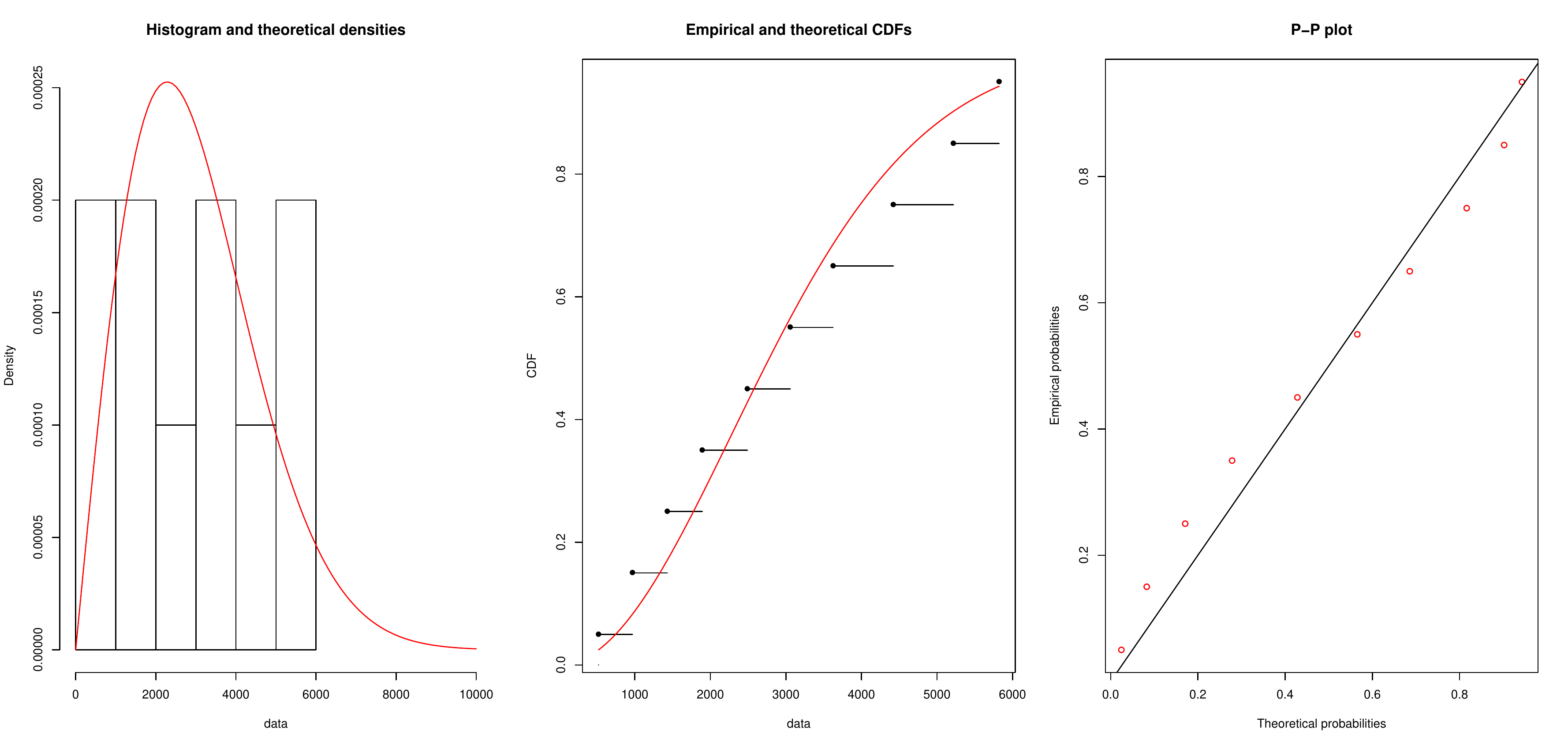}}
\vspace{1cm}
\caption{Data set 1}
\label{fig4}
\end{figure}

\begin{figure}[htbp]
\centering
\resizebox{15cm}{5cm}{\includegraphics[trim=.01cm 2cm .01cm .01cm]{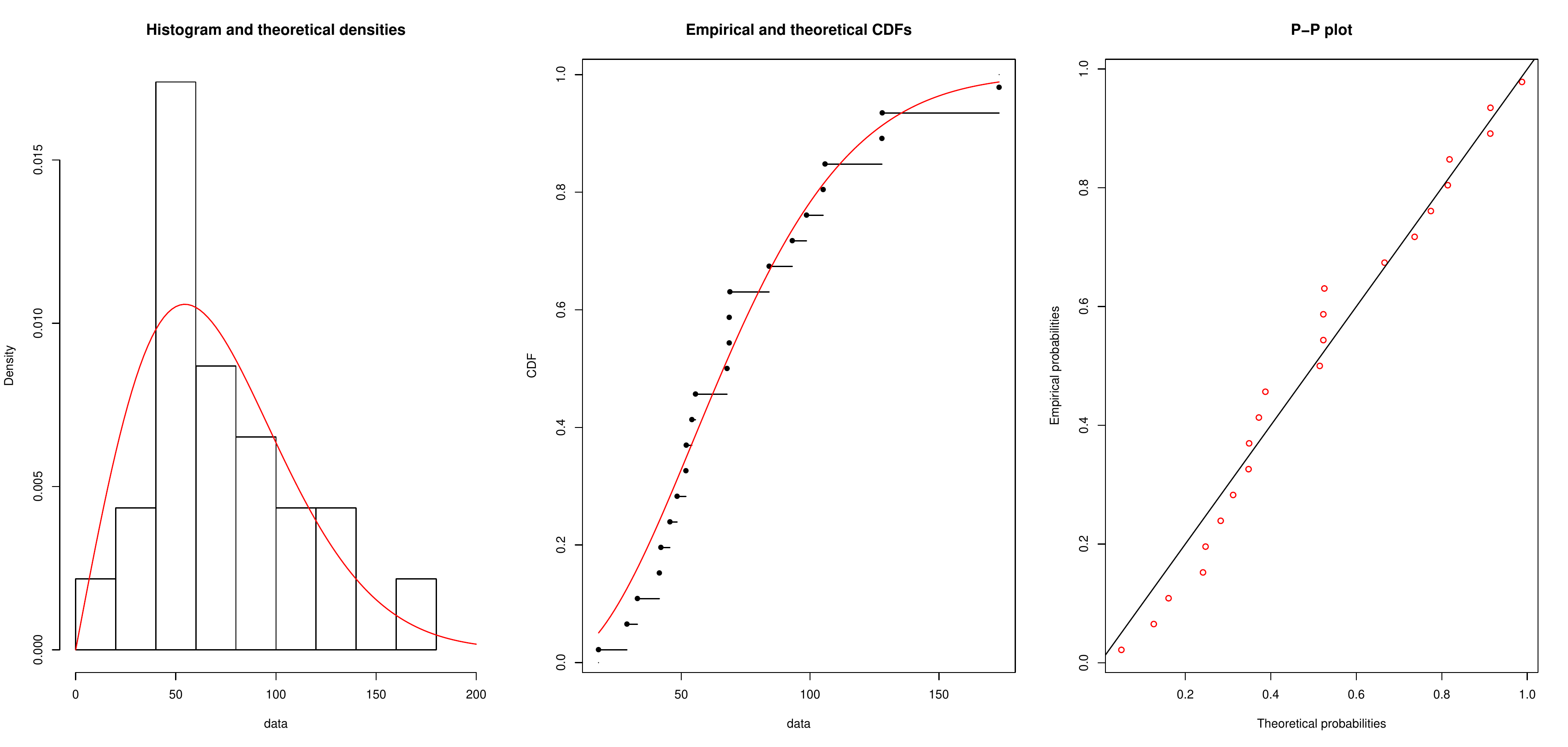}}
\vspace{1cm}
\caption{Data set 2}
\label{fig5}
\end{figure}
%------------------------------------------------------------------------------------------------------------------------------------------------
\begin{table}[ht]
{Table 1: Minimum sample size necessary to ensures that true mean life exceeds given value of $\mu_0$, with probability $P^*$ and the corresponding acceptance number $c$ when $\lambda=0.5$}\\
 \begin{center}
\begin{tiny} 
 \begin{tabular}{|l|l|l|l|l|l|l|l|l|l|l|l|l||l|}\hline
 \multicolumn{1}{|c|}{}&
 \multicolumn{1}{|c|}{}&
 \multicolumn{8}{|c|}{$t/\sigma_0$}\\
 \cline{3-10}
$P^*$ & $c$  & $0.628$ & $0.942$ & $1.257$& $1.571$ & $2.356$ & $3.141$ & $3.927$ & $4.712$  \\
\hline
$0.75$ & $0$  & $5$ & $3$ & $2$& $1$ & $1$ & $1$ & $1$ & $1$  \\
\hline
$$ & $1$  & $9$ & $5$ & $3$& $3$ & $2$ & $2$ & $2$ & $2$  \\
\hline
$$ & $2$  & $12$& $7$ & $5$& $4$ & $3$ & $3$ & $3$ & $3$  \\
\hline
$$ & $3$  & $14$& $8$ & $6$& $5$ & $4$ & $4$ & $4$ & $4$  \\
\hline
$$ & $4$  & $15$& $10$& $8$& $6$ & $5$ & $5$ & $5$ & $5$  \\
\hline
$$ & $5$  & $19$& $11$& $9$& $8$ & $6$ & $6$ & $6$ & $6$  \\
\hline
$$ & $6$  & $23$& $12$& $10$& $9$& $7$ & $7$ & $7$ & $7$  \\
\hline
$$ & $7$  & $28$& $14$& $11$& $10$& $8$& $8$ & $8$ & $8$  \\
\hline
$$ & $8$  & $31$& $16$& $12$& $11$& $9$& $9$ & $9$ & $9$  \\
\hline
$$ & $9$  & $35$& $19$& $13$& $12$& $10$& $10$& $10$& $10$\\
\hline
$$ & $10$ & $39$& $21$& $14$& $13$& $12$& $11$& $11$& $11$\\
\hline
\hline
$0.90$ & $0$  & $8$ & $4$ & $3$& $2$ & $1$ & $1$ & $1$ & $1$\\
\hline
$$ & $1$  & $13$ & $7$ & $4$& $3$ & $2$ & $2$ & $2$ & $2$  \\
\hline
$$ & $2$  & $18$& $9$ & $6$& $5$ & $3$ & $3$ & $3$ & $3$  \\
\hline
$$ & $3$  & $22$& $12$ & $8$& $6$ & $5$ & $4$ & $4$ & $4$  \\
\hline
$$ & $4$  & $27$& $14$& $9$& $8$ & $6$ & $5$ & $5$ & $5$  \\
\hline
$$ & $5$  & $31$& $16$& $11$& $9$ & $7$ & $6$ & $6$ & $6$  \\
\hline
$$ & $6$  & $35$& $18$& $13$& $10$& $8$ & $7$ & $7$ & $7$  \\
\hline
$$ & $7$  & $38$& $20$& $14$& $11$& $9$& $8$ & $8$ & $8$  \\
\hline
$$ & $8$  & $42$& $23$& $16$& $13$& $10$& $9$ & $9$ & $9$  \\
\hline
$$ & $9$  & $46$& $25$& $17$& $14$& $11$& $10$& $10$& $10$\\
\hline
$$ & $10$ & $50$& $27$& $19$& $15$& $12$& $11$& $11$& $11$\\
\hline
\hline
$0.95$ & $0$  & $11$ & $5$ & $3$& $2$ & $1$ & $1$ & $1$ & $1$\\
\hline
$$ & $1$  & $17$ & $8$ & $5$& $4$ & $3$ & $2$ & $2$ & $2$  \\
\hline
$$ & $2$  & $22$& $11$ & $7$& $5$ & $4$ & $3$ & $3$ & $3$  \\
\hline
$$ & $3$  & $27$& $13$ & $9$& $7$ & $5$ & $4$ & $4$ & $4$  \\
\hline
$$ & $4$  & $31$& $16$& $11$& $8$ & $6$ & $5$ & $5$ & $5$  \\
\hline
$$ & $5$  & $36$& $18$& $12$& $10$ & $7$ & $6$ & $6$ & $6$  \\
\hline
$$ & $6$  & $41$& $21$& $14$& $11$& $8$ & $7$ & $7$ & $7$  \\
\hline
$$ & $7$  & $45$& $23$& $16$& $12$& $9$& $8$ & $8$ & $8$  \\
\hline
$$ & $8$  & $49$& $26$& $17$& $14$& $10$& $9$ & $9$ & $9$  \\
\hline
$$ & $9$  & $54$& $28$& $19$& $15$& $11$& $10$& $10$& $10$\\
\hline
$$ & $10$ & $58$& $30$& $21$& $16$& $13$& $11$& $11$& $11$\\
\hline
\hline
$0.99$ & $0$  & $16$ & $8$ & $5$& $3$ & $2$ & $1$ & $1$ & $1$\\
\hline
$$ & $1$  & $23$ & $11$ & $7$& $5$ & $3$ & $2$ & $2$ & $2$  \\
\hline
$$ & $2$  & $30$& $14$ & $9$& $7$ & $4$ & $4$ & $4$ & $4$  \\
\hline
$$ & $3$  & $35$& $17$ & $11$& $8$ & $5$ & $5$ & $4$ & $4$  \\
\hline
$$ & $4$  & $41$& $20$& $13$& $10$ & $7$ & $6$ & $5$ & $5$  \\
\hline
$$ & $5$  & $46$& $23$& $15$& $11$ & $8$ & $7$ & $6$ & $6$  \\
\hline
$$ & $6$  & $51$& $26$& $17$& $13$& $9$ & $8$ & $7$ & $7$  \\
\hline
$$ & $7$  & $56$& $28$& $18$& $14$& $10$& $9$ & $8$ & $8$  \\
\hline
$$ & $8$  & $61$& $31$& $20$& $16$& $11$& $10$ & $9$ & $9$  \\
\hline
$$ & $9$  & $66$& $33$& $22$& $17$& $12$& $11$& $10$& $10$\\
\hline
$$ & $10$ & $71$& $36$& $24$& $18$& $13$& $12$& $11$& $11$\\
\hline
\end{tabular}
\end{tiny}
\end{center}
\label{tab1}
\end{table}

\begin{table}[ht]
{Table 2: OC values for the time truncated acceptance sampling plan $(n,c,\frac{t}{\sigma_0})$, for given $P^*$ when corresponding acceptance number $c=2$, $\lambda=0.5$}\\
 \begin{center}
\begin{tiny} 
 \begin{tabular}{|l|l|l|l|l|l|l|l|l|l|l|l|l||l|}\hline
 \multicolumn{1}{|c|}{}&
 \multicolumn{1}{|c|}{}&
  \multicolumn{1}{|c|}{}&
 \multicolumn{6}{|c|}{$\sigma/\sigma_0$}\\
 \cline{4-9}
$P^*$ & $n$ & $t/\sigma_0$ & $2$ & $4$ & $6$& $8$ & $10$ & $12$\\
\hline
$0.75$ & $12$  & $0.628$ & $0.9515095$ & $0.9988092$& $0.9998861$ & $0.9999791$ & $0.9999944$ & $0.9999981$\\
\hline
$$ & $7$  & $0.942$ & $0.9237801$ & $0.9979221$& $0.9997971$ & $0.9999625$ & $0.99999$ & $0.9999966$  \\
\hline
$$ & $5$  & $1.257$& $0.8935343$ & $0.9967775$& $0.9996784$ & $0.9999401$ & $0.999984$ & $0.9999946$   \\
\hline
$$ & $4$  & $1.571$& $0.8608107$ & $0.9952958$& $0.9995194$ & $0.9999097$ & $0.9999757$ & $0.9999918$  \\
\hline
$$ & $3$  & $2.356$& $0.7554522$& $0.988427$& $0.9987212$ & $0.9997528$ & $0.9999327$ & $0.999977$  \\
\hline
$$ & $3$  & $3.141$& $0.4648425$& $0.9522578$& $0.9937827$ & $0.998722$ & $0.9996415$ & $0.9998754$  \\
\hline
$$ & $3$  & $3.927$& $0.2297293$& $0.8746346$& $0.9801821$& $0.995604$ & $0.9987206$ & $0.9995462$  \\
\hline
$$ & $3$  & $4.712$& $0.09606505$& $0.7554522$& $0.9522344$& $0.988427$& $0.988427$ & $0.9987212$  \\
\hline
\hline
$0.90$ & $18$  & $0.628$ & $0.8684592$ & $0.9959309$& $0.9995928$ & $0.9999241$ & $0.9999797$ & $0.9999931$ \\
\hline
$$ & $9$  & $0.942$ & $0.854900$ & $0.9953096$& $0.9995262$ & $0.9999114$ & $0.9999762$ & $0.9999919$  \\
\hline
$$ & $6$  & $1.257$& $0.8261495$ & $0.993896$& $0.9993724$ & $0.9998819$ & $0.9999682$ & $0.9999892$   \\
\hline
$$ & $5$  & $1.571$& $0.7443875$ & $0.9891916$& $0.9988433$ & $0.9997791$ & $0.9999402$ & $0.9999796$  \\
\hline
$$ & $3$  & $2.356$& $0.7554522$& $0.988427$& $0.9987212$ & $0.9997528$ & $0.9999327$ & $0.999977$  \\
\hline
$$ & $3$  & $3.141$& $0.4648425$& $0.9522578$& $0.9937827$ & $0.998722$ & $0.9996415$ & $0.9998754$  \\
\hline
$$ & $3$  & $3.927$& $0.2297293$& $0.8746346$& $0.9801821$& $0.995604$ & $0.9987206$ & $0.9995462$  \\
\hline
$$ & $3$  & $4.712$& $0.09606505$& $0.7554522$& $0.9522344$& $0.988427$& $0.988427$ & $0.9987212$  \\
\hline
\hline
$0.95$ & $22$  & $0.628$ & $0.797498$ & $0.9927272$& $0.9992501$ & $0.9998587$ & $0.999962$ & $0.9999871$\\
\hline
$$ & $11$  & $0.942$ & $0.7727992$ & $0.9913324$& $0.9990946$ & $0.9998286$ & $0.9999538$ & $0.9999843$  \\
\hline
$$ & $7$  & $1.257$& $0.7506223$ & $0.9898812$& $0.9989282$ & $0.9997961$ & $0.9999449$ & $0.9999812$   \\
\hline
$$ & $5$  & $1.571$& $0.7443875$ & $0.9891916$& $0.9988433$ & $0.9997791$ & $0.9999402$ & $0.9999796$  \\
\hline
$$ & $4$  & $2.356$& $0.4805908$& $0.9615614$& $0.9953012$ & $0.9990579$ & $0.9997389$ & $0.9999098$  \\
\hline
$$ & $3$  & $3.141$& $0.4648425$& $0.9522578$& $0.9937827$ & $0.998722$ & $0.9996415$ & $0.9998754$  \\
\hline
$$ & $3$  & $3.927$& $0.2297293$& $0.8746346$& $0.9801821$& $0.995604$ & $0.9987206$ & $0.9995462$  \\
\hline
$$ & $3$  & $4.712$& $0.09606505$& $0.7554522$& $0.9522344$& $0.988427$& $0.988427$ & $0.9987212$  \\
\hline
\hline
$0.99$ & $30$  & $0.628$ & $0.6403632$ & $0.9827918$& $0.998117$ & $0.9996377$ & $0.9999015$ & $0.9999664$\\
\hline
$$ & $14$  & $0.942$ & $0.6397935$ & $0.982543$& $0.998083$ & $0.9996306$ & $0.9998995$ & $0.9999657$  \\
\hline
$$ & $9$  & $1.257$& $0.5928948$ & $0.9781937$& $0.99755$ & $0.999524$ & $0.9998700$ & $0.9999555$   \\
\hline
$$ & $7$  & $1.571$& $0.5026545$ & $0.9679963$& $0.9962476$ & $0.9992595$ & $0.9997963$ & $0.9999300$  \\
\hline
$$ & $4$  & $2.356$& $0.4805908$& $0.9615614$& $0.9953012$ & $0.9990579$ & $0.9997389$ & $0.9999098$  \\
\hline
$$ & $4$  & $3.141$& $0.1628271$& $0.8609899$& $0.9785604$ & $0.995304$ & $0.9986423$ & $0.9995202$  \\
\hline
$$ & $3$  & $3.927$& $0.2297293$& $0.8746346$& $0.9801821$& $0.995604$ & $0.9987206$ & $0.9995462$  \\
\hline
$$ & $3$  & $4.712$& $0.09606505$& $0.7554522$& $0.9522344$& $0.988427$& $0.988427$ & $0.9987212$  \\
\hline
\end{tabular}
\end{tiny}
\end{center}
\label{tab2}
\end{table}

\begin{table}[ht]
{Table 3: Minimum ratio of true mean life to specified mean life for the acceptance of lot with producer risk $0.05$, when $\lambda=0.5$}\\
 \begin{center}
\begin{tiny} 
 \begin{tabular}{|l|l|l|l|l|l|l|l|l|l|l|l|l||l|}\hline
 \multicolumn{1}{|c|}{}&
 \multicolumn{1}{|c|}{}&
 \multicolumn{8}{|c|}{$t/\sigma_0$}\\
 \cline{3-10}
$P^*$ & $c$  & $0.628$ & $0.942$ & $1.257$& $1.571$ & $2.356$ & $3.141$ & $3.927$ & $4.712$  \\
\hline
$0.75$ & $0$  & $5.37$ & $6.24$ & $6.8$& $6$ & $9$ & $12$ & $15$ & $18$  \\
\hline
$$ & $1$  & $2.66$ & $2.89$ & $2.85$& $3.56$ & $4.03$ & $5.37$ & $6.72$ & $8.06$  \\
\hline
$$ & $2$  & $1.99$& $2.19$ & $2.37$& $2.53$ & $2.98$ & $3.97$ & $4.96$ & $5.95$  \\
\hline
$$ & $3$  & $1.64$& $1.76$ & $1.92$& $2.08$ & $2.51$ & $3.34$ & $4.18$ & $5.01$  \\
\hline
$$ & $4$  & $1.39$& $1.61$& $1.85$& $1.82$ & $2.24$ & $2.98$ & $3.73$ & $4.47$  \\
\hline
$$ & $5$  & $1.36$& $1.44$& $1.66$& $1.88$ & $2.06$ & $2.74$ & $3.43$ & $4.11$  \\
\hline
$$ & $6$  & $1.34$& $1.32$& $1.52$& $1.73$& $1.93$ & $2.57$ & $3.21$ & $3.85$  \\
\hline
$$ & $7$  & $1.34$& $1.29$& $1.42$& $1.62$& $1.83$& $2.44$ & $3.05$ & $3.66$  \\
\hline
$$ & $8$  & $1.3$& $1.27$& $1.34$& $1.54$& $1.75$& $2.34$ & $2.92$ & $3.5$  \\
\hline
$$ & $9$  & $1.28$& $1.3$& $1.28$& $1.47$& $1.69$& $2.25$& $2.81$& $3.37$\\
\hline
$$ & $10$ & $1.27$& $1.28$& $1.22$& $1.41$& $1.9$& $2.18$& $2.72$& $3.27$\\
\hline
\hline
$0.90$ & $0$  & $7.97$ & $8.06$ & $8.33$& $8.49$ & $9$ & $12$ & $15$ & $18$\\
\hline
$$ & $1$  & $3.23$ & $3.48$ & $3.39$& $3.56$ & $4.03$ & $5.37$ & $6.72$ & $8.06$  \\
\hline
$$ & $2$  & $2.47$& $2.54$ & $2.66$& $2.96$ & $2.98$ & $3.97$ & $4.96$ & $5.95$  \\
\hline
$$ & $3$  & $2.11$& $2.25$ & $2.34$& $2.4$ & $3.11$ & $3.34$ & $4.18$ & $5.01$  \\
\hline
$$ & $4$  & $1.94$& $2$& $2.01$& $2.31$ & $2.73$ & $2.98$ & $3.73$ & $4.47$  \\
\hline
$$ & $5$  & $1.79$& $1.84$& $1.92$& $2.07$ & $3.31$ & $2.74$ & $3.43$ & $4.11$  \\
\hline
$$ & $6$  & $1.7$& $1.73$& $1.86$& $1.9$& $2.3$ & $2.57$ & $3.21$ & $3.85$  \\
\hline
$$ & $7$  & $1.6$& $1.64$& $1.72$& $1.77$& $1.45$& $2.44$ & $3.05$ & $3.66$  \\
\hline
$$ & $8$  & $1.54$& $1.62$& $1.69$& $1.8$& $2.06$& $2.34$ & $2.92$ & $3.5$  \\
\hline
$$ & $9$  & $1.5$& $1.57$& $1.61$& $1.71$& $1.98$& $2.25$& $2.81$& $3.37$\\
\hline
$$ & $10$ & $1.47$& $1.52$& $1.58$& $1.63$& $1.9$& $2.18$& $2.72$& $3.27$\\
\hline
\hline
$0.95$ & $0$  & $7.97$ & $8.06$ & $8.33$& $8.49$ & $9$ & $12$ & $15$ & $18$\\
\hline
$$ & $1$  & $3.71$ & $3.74$ & $3.86$& $4.24$ & $5.33$ & $5.37$ & $6.72$ & $8.06$  \\
\hline
$$ & $2$  & $2.76$& $2.85$ & $2.93$& $2.96$ & $3.79$ & $3.97$ & $4.96$ & $5.95$  \\
\hline
$$ & $3$  & $2.35$& $2.36$ & $2.52$& $2.68$ & $3.12$ & $3.34$ & $4.18$ & $5.01$  \\
\hline
$$ & $4$  & $2.09$& $2.16$& $2.3$& $2.31$ & $2.73$ & $2.98$ & $3.73$ & $4.47$  \\
\hline
$$ & $5$  & $1.95$& $1.97$& $2.04$& $2.24$ & $2.48$ &$2.74$ & $3.43$ & $4.11$  \\
\hline
$$ & $6$  & $1.85$& $1.9$& $1.96$& $2.05$& $2.3$ & $2.57$ & $3.21$ & $3.85$  \\
\hline
$$ & $7$  & $1.75$& $1.79$& $1.89$& $1.91$& $2.17$& $2.44$ & $3.05$ & $3.66$  \\
\hline
$$ & $8$  & $1.68$& $1.75$& $1.77$& $1.91$& $2.06$& $2.34$ & $2.92$ & $3.5$  \\
\hline
$$ & $9$  & $1.64$& $1.68$& $1.74$& $1.81$& $1.98$& $2.25$& $2.81$& $3.37$\\
\hline
$$ & $10$ & $1.59$& $1.63$& $1.71$& $1.73$& $2.11$& $2.18$& $2.72$& $3.27$\\
\hline
\hline
$0.99$ & $0$  & $9.61$ & $10.19$ & $10.75$& $10.4$ & $12.74$ & $12$ & $15$ & $18$\\
\hline
$$ & $1$  & $4.33$ & $4.43$ & $4.64$& $4.82$ & $5.33$ & $5.37$ & $6.72$ & $8.06$  \\
\hline
$$ & $2$  & $3.24$& $3.24$ & $3.39$& $3.66$ & $3.79$ & $5.05$ & $4.96$ & $5.95$  \\
\hline
$$ & $3$  & $2.69$& $2.74$ & $2.85$& $2.93$ & $3.11$ & $4.14$ & $4.18$ & $5.01$  \\
\hline
$$ & $4$  & $2.42$& $2.46$& $2.55$& $2.7$ & $3.13$ & $3.64$ & $3.73$ & $4.47$  \\
\hline
$$ & $5$  & $2.22$& $2.28$& $2.36$& $2.4$ & $2.82$ & $3.31$ & $3.43$ & $4.11$  \\
\hline
$$ & $6$  & $2.08$& $2.15$& $2.22$& $2.32$& $2.6$ & $3.07$ & $3.21$ & $3.85$  \\
\hline
$$ & $7$  & $1.97$& $2.07$& $2.05$& $2.15$& $2.43$& $2.89$ & $3.05$ & $3.66$  \\
\hline
$$ & $8$  & $1.89$& $1.95$& $1.98$& $2.12$& $2.3$& $2.75$ & $2.92$ & $3.5$  \\
\hline
$$ & $9$  & $1.83$& $1.86$& $1.92$& $2$& $2.2$& $2.63$& $2.81$& $3.37$\\
\hline
$$ & $10$ & $1.78$& $1.82$& $1.88$& $1.9$& $2.11$& $2.54$& $2.72$& $3.27$\\
\hline
\end{tabular}
\end{tiny}
\end{center}
\label{tab1}
\end{table}

\begin{table}[ht]
{Table 4: Producer's risks when $P^*=0.95$, $c=2$ and $\lambda=0.5$  }\\
 \begin{center}
 \begin{tiny}
 \begin{tabular}{|l|l|l|l|l|l|l|l|l|l|l|l|l||l|}\hline
 \multicolumn{1}{|c|}{}&
 \multicolumn{1}{|c|}{}&
  \multicolumn{1}{|c|}{}&
 \multicolumn{6}{|c|}{$\sigma/\sigma_0$}\\
 \cline{4-9}
$P^*$ & $n$  & $t/\sigma_0$ & $2$ & $4$& $6$ & $8$ & $10$ & $12$  \\
\hline
$0.75$ & $12$  & $0.628$ & $0.0484904$ & $0.00119081$& $0.0001138951$ & $2.089087e-05$ & $5.553451e-06$ & $1.874017e-06$\\
\hline
$$ & $7$  & $0.942$ & $0.0762198$ & $0.00207792$& $0.0002029209$ & $3.749623e-05$ & $1.000202e-05$ & $3.38152e-06$  \\
\hline
$$ & $5$  & $1.257$& $0.1064657$ & $0.00322254$& $0.0003215574$ & $5.987798e-05$ & $1.602984e-05$ & $5.43007e-06$   \\
\hline
$$ & $4$  & $1.571$& $0.1391893$ & $0.00470420$& $0.0004806372$ & $9.026695e-05$ & $2.426185e-05$ & $8.236572e-06$  \\
\hline
$$ & $3$  & $2.356$& $0.2445478$& $0.01157296$& $0.0012787930$ & $0.0002471642$ & $6.733724e-05$ & $2.303011e-05$  \\
\hline
$$ & $3$  & $3.141$& $0.5351575$& $0.04774217$& $0.0062173340$ & $0.0012780300$ & $0.0003585210$ & $0.000124607$  \\
\hline
$$ & $3$  & $3.927$& $0.7702707$& $0.12536540$& $0.0198179100$& $0.0043960330$ & $0.0012794040$ & $0.000453820$  \\
\hline
$$ & $3$  & $4.712$& $0.9039350$& $0.24454780$& $0.0477656100$& $0.0115729600$& $0.0035181410$ & $0.001278790$  \\
\hline
\hline
$0.90$ & $18$  & $0.628$ & $0.1315408$ & $0.00406908$& $0.0004072074$ & $7.589678e-05$ & $2.03265e-05$ & $ 6.887046e-06$\\
\hline
$$ & $9$  & $0.942$ & $0.1451000$ & $0.00469037$& $0.0004737901$ & $8.860401e-05$ & $2.376705e-05$ & $8.059686e-06$  \\
\hline
$$ & $6$  & $1.257$& $0.1738505$ & $0.00610397$& $0.0006275758$ & $0.0001181129$ & $3.177695e-05$ & $1.079347e-05$   \\
\hline
$$ & $5$  & $1.571$& $0.2556125$ & $0.01080841$& $0.0011566700$ & $0.000220857$ & $5.982195e-05$ & $2.039445e-05$  \\
\hline
$$ & $3$  & $2.356$& $0.2445478$& $0.01157296$& $0.0012787930$ & $0.0002471642$ & $6.733724e-05$ & $2.303011e-05$  \\
\hline
$$ & $3$  & $3.141$& $0.5351575$& $0.04774217$& $0.0062173340$ & $0.0012780300$ & $0.0003585210$ & $0.000124607$  \\
\hline
$$ & $3$  & $3.927$& $0.7702707$& $0.12536540$& $0.0198179100$& $0.0043960330$ & $0.0012794040$ & $0.000453820$  \\
\hline
$$ & $3$  & $4.712$& $0.9039350$& $0.24454780$& $0.0477656100$& $0.0115729600$& $0.0035181410$ & $0.001278790$  \\
\hline
\hline
$0.95$ & $22$  & $0.628$ & $0.2025020$ & $0.00727277$& $0.000749947$ & $0.0001412735$ & $3.8023e-05$ & $1.2917e-05$\\
\hline
$$ & $11$  & $0.942$ & $0.2272008$ & $0.00866759$& $0.000905443$ & $0.0001713640$ & $4.6223e-05$ & $1.5722e-05$  \\
\hline
$$ & $7$  & $1.257$& $0.2493777$ & $0.01011879$& $0.001071766$ & $0.0002038644$ & $5.5119e-05$ & $1.8772e-05$   \\
\hline
$$ & $5$  & $1.571$& $0.2556125$ & $0.01080841$& $0.001156670$ & $0.0002208570$ & $5.9821e-05$ & $2.0394e-05$  \\
\hline
$$ & $4$  & $2.356$& $0.5194092$& $0.03843864$& $0.004698762$ & $0.0009421228$ & $0.0002611300$ & $9.0154e-05$  \\
\hline
$$ & $3$  & $3.141$& $0.5351575$& $0.04774217$& $0.006217334$ & $0.0012780300$ & $0.0003585210$ & $0.000124607$  \\
\hline
$$ & $3$  & $3.927$& $0.7702707$& $0.12536540$& $0.019817910$& $0.0043960330$ & $0.0012794040$ & $0.000453820$  \\
\hline
$$ & $3$  & $4.712$& $0.9039350$& $0.24454780$& $0.047765610$& $0.0115729600$& $0.0035181410$ & $0.001278790$  \\
\hline
\hline
$0.99$ & $30$  & $0.628$ & $0.3596368$ & $0.01720824$& $0.00188304$ & $0.0003623232$ & $9.848695e-05$ & $3.364032e-05$\\
\hline
$$ & $14$  & $0.942$ & $0.3602065$ & $0.01745705$& $0.001917025$ & $0.0003693541$ & $0.0001004617$ & $3.432682e-05$  \\
\hline
$$ & $9$  & $1.257$& $0.4071052$ & $0.02180631$& $0.00244995$ & $0.0004759724$ & $0.0001299673$ & $4.45035e-05$   \\
\hline
$$ & $7$  & $1.571$& $0.4973455$ & $0.03200367$& $0.003752446$ & $0.0007404669$ & $0.0002036755$ & $7.002275e-05$  \\
\hline
$$ & $4$  & $2.356$& $0.5194092$& $0.03843864$& $0.004698762$ & $0.0009421228$ & $0.0002611300$ & $9.0154e-05$  \\
\hline
$$ & $4$  & $3.141$& $0.8371729$& $0.13901010$& $0.0214396$ & $0.00469604$ & $0.001357677$ & $0.0004797575$  \\
\hline
$$ & $3$  & $3.927$& $0.7702707$& $0.12536540$& $0.019817910$& $0.0043960330$ & $0.0012794040$ & $0.000453820$  \\
\hline
$$ & $3$  & $4.712$& $0.9039350$& $0.24454780$& $0.047765610$& $0.0115729600$& $0.0035181410$ & $0.001278790$  \\
\hline
\end{tabular}
\end{tiny}
\end{center}
\label{tab4}
\end{table}

\begin{table}[ht]
{Table 5: Descriptive statistics for the considered data sets}\\
\begin{center}
\begin{tabular}{|l|l|l|l|l|l|l|l|l|l|l|l|l|}
\hline
Data    & Minimum & $Q_1$         &  Median & Mean & $Q_3$         & Maximum & $CS$ & $CK$   \\
set     &         &      &         &      &     &         &    &      \\
\hline
I& $519$ & $1546$ & $2774$ & $2945$  & $4223$  &   $5823$ & $0.2448947$ & $1.800022$     \\
\hline
II& $17.88$  & $47$ & $67.80$ & $72.22$    & $95.88$  &   $173.40$ & $0.9412634$ & $3.486325$     \\
\hline
\end{tabular}
\end{center}
\label{tab5}
\end{table}

\begin{table}[ht]
{Table 6: Model fitting summary for the considered data sets}\\
\begin{center}
\begin{tabular}{|l|l|l|l|l|l|l|l|l|l|l|l|l|}
\hline
Data set& Log likelihood & AIC &  BIC & KS Statistic & KS $p$ value  \\
\hline
I& $-88.12364$ & $180.2473$ & $180.8525$ & $0.12909$  & $0.9884$  \\
\hline
II& $-113.7319$  & $231.4638$ & $233.7348$ & $0.12694$    & $0.8525$  \\
\hline
\end{tabular}
\end{center}
\label{tab6}
\end{table}

\end{document}